# Inhomogeneous ferrimagnetic-like behavior in $Gd_{2/3}Ca_{1/3}MnO_3$ single crystals

N. Haberkorn

Comisión Nacional de Energía Atómica, Centro Atómico Bariloche, S. C. de Bariloche, 8400 R. N., Argentina. and

Instituto Balseiro, Universidad Nacional de Cuyo and Comisión Nacional de Energía Atómica, S. C. de Bariloche, 8400 R. N., Argentina

S. Larrégola

Facultad de Química, Bioquímica y Farmacia, Universidad Nacional de San Luis, Chacabuco y Pedernera, San Luis 5700, Argentina.

D. Franco

Facultad de Ciencias Químicas, Universidad Nacional de Córdoba, 5000 Córdoba, Argentina

G. Nieva

Comisión Nacional de Energía Atómica, Centro Atómico Bariloche, S. C. de Bariloche, 8400 R. N., Argentina. and

Instituto Balseiro, Universidad Nacional de Cuyo and Comisión Nacional de Energía Atómica, S. C. de Bariloche, 8400 R. N., Argentina

We present a study of the magnetic properties of $Gd_{2/3}Ca_{1/3}MnO_3$ single crystals at low temperatures. We show that this material behave as an inhomogeneous ferrimagnet. In addition to small saturation magnetization at 5 K, we have found history dependent effects in the magnetization and the presence of exchange bias. These features are compatible with microscopic phase separation in the clean $Gd_{2/3}Ca_{1/3}MnO_3$ system studied.

nhaberk@cab.cnea.gov.ar    TE / FAX: 542944445171 / 2944445299

Transition metal oxides with perovskite structure have generated great interest due to the rich variety of their electrical and magnetic properties. These include, for example, high temperature superconductivity and colossal magnetorresistance (CMR).[1,2] In hole-doped perovskites $R_{1-x}A_x MnO_3$ ($RA_x$MO R: Rare earths, and A: Ca, Sr) showing CMR, the electrical and magnetic properties are known to be very sensitive to the lattice parameters, the $Mn^{3+}$ / $Mn^{4+}$ ratio and the oxygen content.[1] By modification of these parameters it is possible to obtain complex phase diagrams that include ferromagnetism (FM), antiferromagnetism (AF), weak FM, spin canting and, in some cases, spatial inhomogeneity related to multiphase coexistence.[1,3] The phenomenon of multiphase coexistence or phase separation corresponds to the simultaneous existence of mutually penetrating submicrometer sub-phases with slightly different electronic density giving rise to different magnetic behavior. This electronic phase separation is associated with spatial inhomogeneity which in turn is related to local crystalline distortion. The mismatch between the size of different ions could be expressed by the values of the tolerance factor, $t = \dfrac{\langle r_{R;A}\rangle + r_O}{\sqrt{2}(r_{Mn}+r_O)}$, and size disorder at the R,A-site, $\sigma^2 \sum x_i r_i^2 - \langle r_i \rangle^2$, where $\langle r_{R,A}\rangle$ corresponds to the average size of the R,A-site cations, and, $x_i$ and $r_i$ are the fractional occupancies and ionic radii of the i cations.[4,5] In GdCa$_x$MO for x ≈ 1/3, we can estimate t and $\sigma^2$, 0.891 and 0.0025 A$^2$, respectively, a result quite similar to YCa$_{1/3}$MO (t ≈ 0.884, $\sigma^2$ ≈ 0.0014 A$^2$). Whereas YCa$_{1/3}$MO shows short ferromagnetic order and a spin glass-like (SGL) behavior at low temperatures (T < 30 K),[6] in GdCa$_{1/3}$MO a ferrimagnetic behavior was reported, associated with the antiferromagnetic (AF) order of the Gd and Mn sublattices at low temperatures.[7,8] The presence of a SGL behavior in YCa$_{1/3}$MO could be related to the large local lattice distortion and associated with phase coexistence,[3,6] similar to that found in PrCa$_{1/3}$MO.[9] The GdCa$_{1/3}$MO compound presents ferrimagnetic behavior with Curie temperatures of around 50 and 80 K, for the Gd and Mn sublattices, and a compensation temperature (T$_{comp}$) of ≈ 15 K.[7,8] At T$_{comp}$, the rare-earth and transition metal sublattice magnetizations at zero field exactly cancel.

In this work, we analyze the magnetic behavior at low temperatures of GdCa$_{1/3}$MO single crystals. The samples were grown by the floating zone technique from isostatically pressed and pre-sintered rods of the same nominal composition. The phase purity of the single crystals was probed by x-ray diffraction, and the composition was checked by energy dispersive spectroscopy (EDS). Magnetic properties were measured in a commercial SQUID magnetometer. Curie temperature ($T_c$) was estimated from the inflection point of the field-cooled (FC) magnetization (*M*) versus *T* curves at low applied magnetic fields. In the FC procedure the samples were cooled from 150 K, under an applied field between 0 and 5 T. Cooling (FCC) and warming (FCW) *M* versus *T* measurements were performed. The results that will be presented correspond to the same crystal, being representative of all measured crystals.

Powder x-ray diffraction patterns obtained by grinding several $GdCa_{1/3}MO$ single crystals show single phase orthorhombic Pbnm(n°62) structure. Two crystalline directions were identified in the single crystal used in the magnetic measurements. Figure 1 shows a schematic picture of the crystalline axis and its respective x-ray diffraction pattern. The (020) and (200) orthorhombic reflections are equivalent to the family of (110) reflection expected for the pseudo orthorhombic or pseudo cubic (p-cub) lattice formed by the cations. The lattice parameters are $b/\sqrt{2} \approx 0.393$ nm, $a/\sqrt{2} \approx 0.381$ nm. Taking these into consideration, a face rotated 90° from those is equivalent to the (100) axis (see figure 1).

Figure 2a shows *M* versus *T* curves for field H = 7.5 kOe applied along different crystal orientations of $GdCa_{1/3}MO$ single crystal. The results show an inflection of the magnetization at approximately 80 K associated with the ferromagnetic order of the Mn cations. Below 50 K the magnetization decreases due to the compensation originated by the magnetic order of the Gd sublattice (Gd-Mn interaction). Depending on the applied magnetic field the magnetization goes to negative values (H < 2.5 kOe, not shown) or begins to increase at $T_{comp} \approx 16$ K.[7,8] Figures 2b and 2c show the hysteretic *M* versus *T* behavior around $T_{comp}$ for two different applied magnetic fields in the $(100)_{p-cub}$ axis. The differences in cooling and warming measurements could be associated with a change in the domain size. This fact is also manifested in the coercive field. Hysteresis loops at the same temperature range present different coercive fields when the temperature is reached by cooling or warming (not shown). Hysteresis in magnetization was previously reported in $YCa_{1/3}MO$,[6] and it was associated with a spin glass like (SGL) behavior. In this case, as in our experiments a dynamic phase coexistence could be present since long local distortions are present in both materials.[3] As we will show later, a possible phase separation is also supported by two different facts: the low $M_s$ at low temperatures, and the presence of exchange bias (*EB*) near $T_{comp}$. Different curves in figure 2a make evident the crystalline anisotropy effect. This anisotropy is also manifested in the hysteresis loops presented in figure 3. At 60 K it is easier to magnetize the Mn along the (020) direction than along $(100)_{p-cub}$ axis (see figure 3a).

While at the lowest measured temperatures, where the Gd influences the magnetization, the easier axis corresponds to the $(100)_{p\text{-cub}}$ axis(see figures 3b and 3c). Although more studies are necessary to clarify this point, the anisotropy difference of the Mn and Gd sublattices could produce canting between them.

Figure 4 shows the saturation magnetization ($M_s$) obtained from hysteresis magnetic loops like those shown in figure 3. In the $M_s$ estimated a paramagnetic signal was subtracted. The possible phase coexistence is supported by the low $M_s$ value at 5 K ≈ 80 emu / cm$^3$. This value is approximately half of the expected value considering ferrimagnetic order. The saturation magnetization per mole of GdCa$_{1/3}$MO expected from the Gd$^{3+}$ s = 7/2, l = 0 is µ = (2/3 x 2 x 7/2) µB = 4.67 µB, while high-spin manganese gives spin only, µ = gsµB, g = 2 so µ = 2 µB[2/3x2 (for Mn$^{3+}$)+ 1/3x3/2 (from Mn$^{4+}$)]= 3.67 µB.[7] Considering these magnetizations, we expect a $M_s$ ≈ 1.00 µB ≈ 160 emu/cm$^3$.[7] Magnetic hysteresis loops also show a high paramagnetic like signal at different temperatures (see figure 3). Although at low temperatures it could be associated with sublattice rotation due to canting,[7] it could also be associated with phase coexistence. It is important to remark that the contribution of paramagnetic Gd moment alone can not explain the high paramagnetic signal, because in this case a high $M_s$ should be expected from the non compensated ferromagnetic Mn moments. Figure 5 shows the temperature dependence of the coercive field,

$H_c$ =| $H_{c1}$−$H_{c2}$ | /2, where $H_{c1}$ and $H_{c2}$ are the fields for zero magnetization at both branches of the hysteresis loops for field excursions up to 1 and 3 T. The $H_c$ temperature dependence shows a non monotonic decrease when the temperature is raised, resulting quite different to the continuous and smooth decrease measured by O. Peña et al.[8] We observe a drop of $H_c$ around Tcomp, which is a consequence of the superposition of two signals: a ferrimagnetic one, responsible for the loops, and a paramagnetic one.[10] As we discuss previously, the paramagnetic like behavior could be a consequence of weakly coupled sublattices,[7] or phase coexistence. The coercive field also makes evident the crystalline anisotropy (see figure 5). We observe that the differences in $H_c$ for the $(100)_{p\text{-cub}}$ and (020) axis are more important at temperatures lower than 30 K, in this temperature range the Gd sublattice magnetization starts to play a more important role. However, in our case, considering a possible phase coexistence, the shape anisotropy of the small domains could be affecting the $H_c$ values.[11] Another feature in the hysteresis loops (see inset figure 3b) is the shift with respect to zero field of the magnetization, *i. e*. exchange bias (EB). The presence of *EB*, associated to the presence of FM / AF magnetic interfaces is usually found in artificially designed materials like FM / AF multilayers or more

recently, in manganite with phase coexistence.[12, 13] Materials with AFM / ferrimagnetic and FM / ferrimagnetic interfaces also could show EB.[14] The magnitude of the effect depends on a number of parameters including AF and FM domain size, interfacial roughness, AF and FM anisotropy, etc.

Figure 6 shows the T dependence of the *EB* field ($H_{eb} = |H_{c1} + H_{c2}|/2$), obtained from hysteresis loops at different FC field, H = 1 and 3 T, in two crystalline directions. The exchange bias fields show a sharp increase and change sign around $T_{comp}$. This behavior is similar to that found in ferrimagnetic / FM multilayer.[10, 15] Since we are measuring a single crystalline sample we expect a homogeneous ferrimagnet due to the absence of interfaces. However the presence of EB should be associated in our case with a very small domain size distribution with different $T_{comp}$. This is supported by the suppression of the Heb when the magnetic fields excursions in the loops are increased. For example, loops in a range -1 T < H < 1 T show Heb several times higher than loops in a range -3 T < H < 3 T, and the EB effect disappear for loops in a range -5 T < H < 5 T. The change of sign in $H_{eb}$ near of $T_{comp}$ is a consequence of domain rotation when the Gd sublattices dominates the magnetization.

In summary, we studied the magnetic properties of GdCa1/3MO single crystals. This cleans system, highly locally distorted, shows characteristics typically found in inhomogeneous ferrimagnets. Several features are compatible with phase coexistence in the single crystals: the Ms at 5 K is approximately half of the expected value considering a ferrimagnetic order; hysteric behavior on cooling and warming *M* versus *T* curves; and the presence of exchange bias near $T_{comp}$.

## Acknowledgments

This work was partially supported by CONICET PIP5251 and ANPCYT PICT PICT00-03- 08937. N. H. and G. N. are member of CONICET.

.

Figure 1. X-ray diffraction patterns for different crystalline axis in the studied $GdCa_{1/3}MO$ single crystal. Arrows indicate the equivalent crystalline orientations in the pseudo cubic structure given by the cations sublattices.

Figure 2. (a) Magnetization ($M$) versus Temperature ($T$) at 7500 Oe for different crystalline axis in a $GdCa_{1/3}MO$ single crystal. Open Circle: (020); Close square: (200); and, Open triangle: $(100)_{p\text{-cub}}$. (b) and (c) Magnetization ($M$) versus Temperature ($T$) in the $(100)_{p\text{-cub}}$ axis at 2500 and 5000 Oe, respectively. Close circle: cooling; Open circle: warming.

Figure. 3: Magnetization ($M$) versus magnetic field ($H$) at different temperatures: (a) 60 K; (b) 30 K; and (c) 5 K. The inset in (b) shows the presence of exchange bias at 18 K for a magnetic loop in a magnetic field range -1 T < H < 1 T.

Figure 4. Saturation magnetization ($M_S$) versus Temperature ($T$) obtained from magnetic hysteresis loops in a $GdCa_{1/3}MO$ single crystal. Dashed line is guide by to the eye.

Figure 5. Coercive field ($H_c$) vs Temperature ($T$) for different crystalline axis in a $GdCa_{1/3}MO$ single crystal.

Figure 6. Exchange bias field ($H_{eb}$) vs Temperature ($T$) for different crystalline axis in a $GdCa_{1/3}MO$ single crystal.

**Figure**

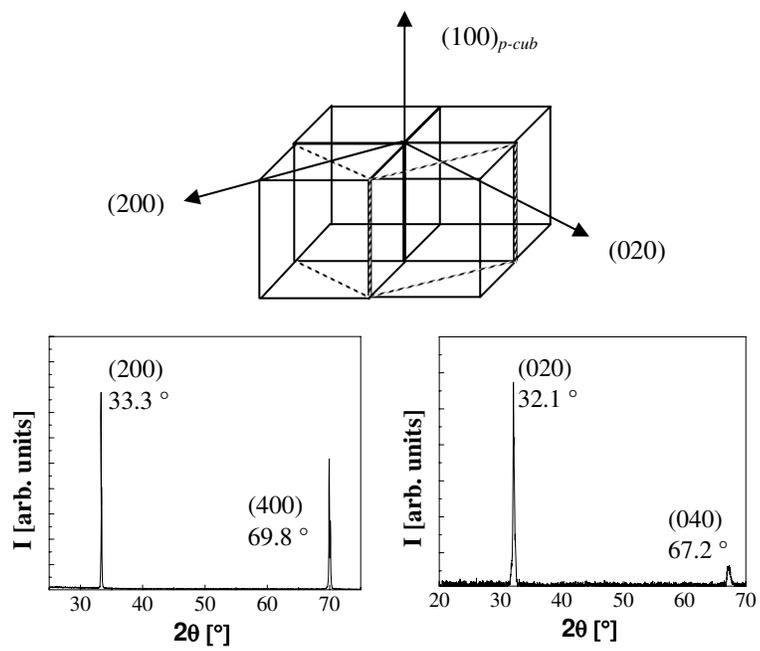

Figure

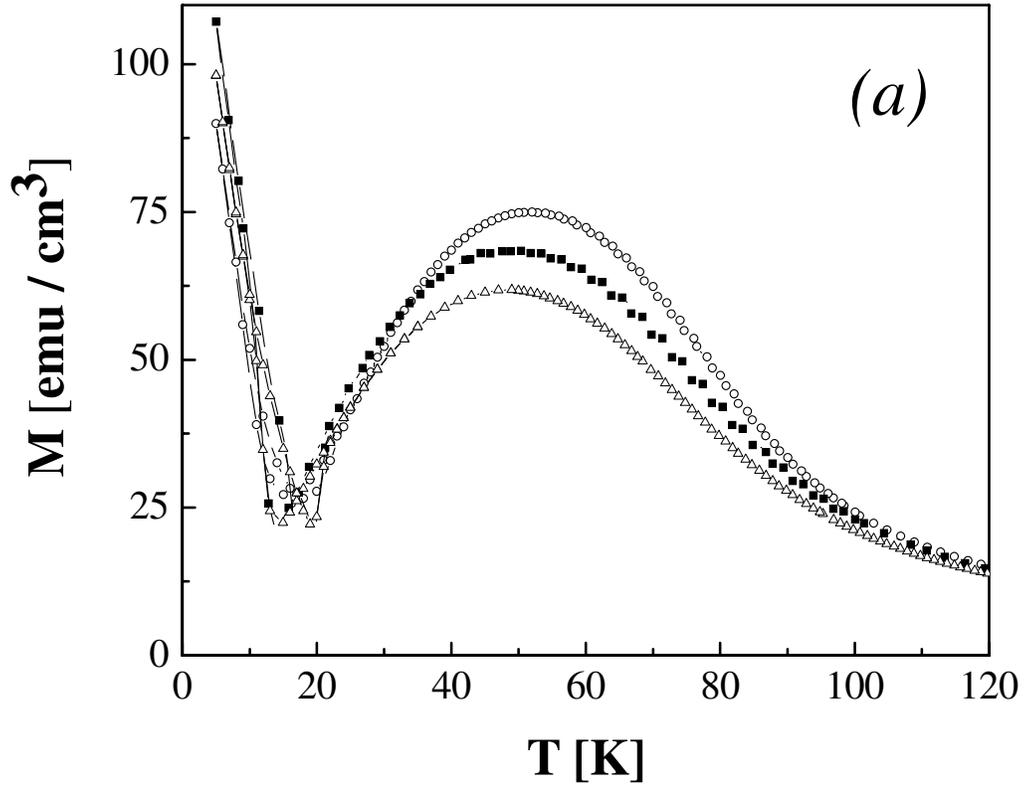
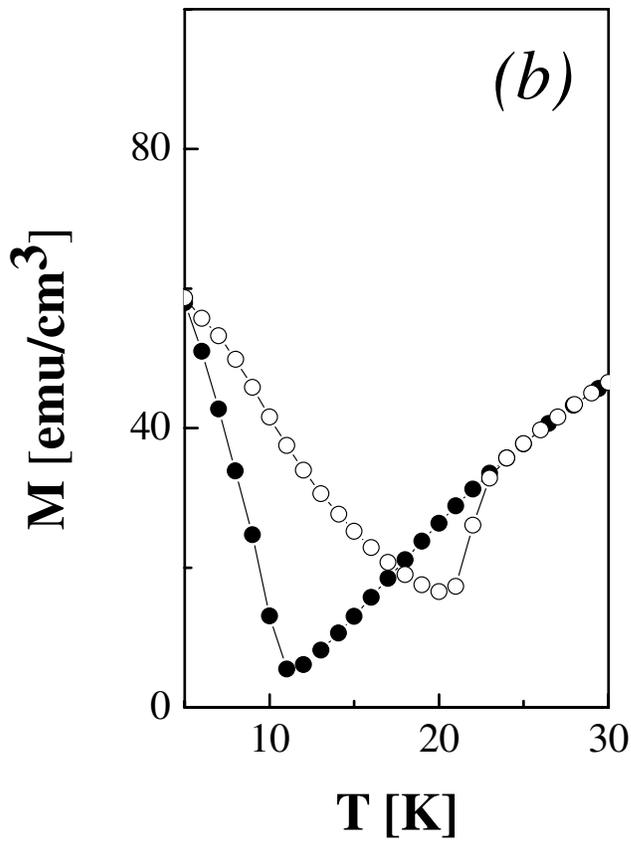
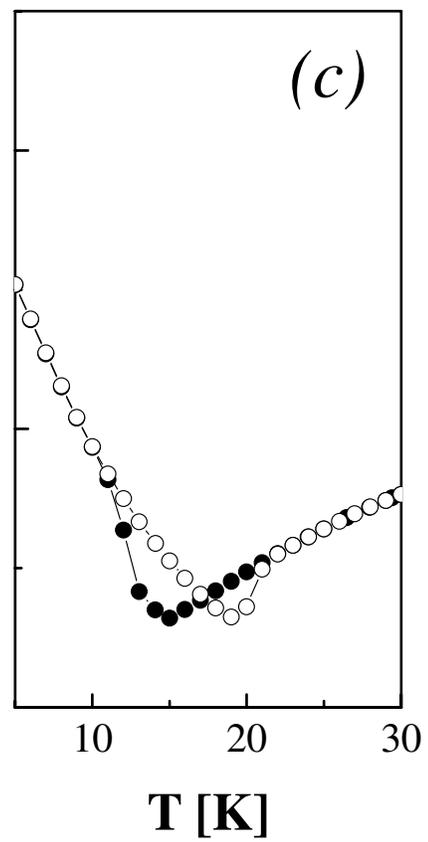

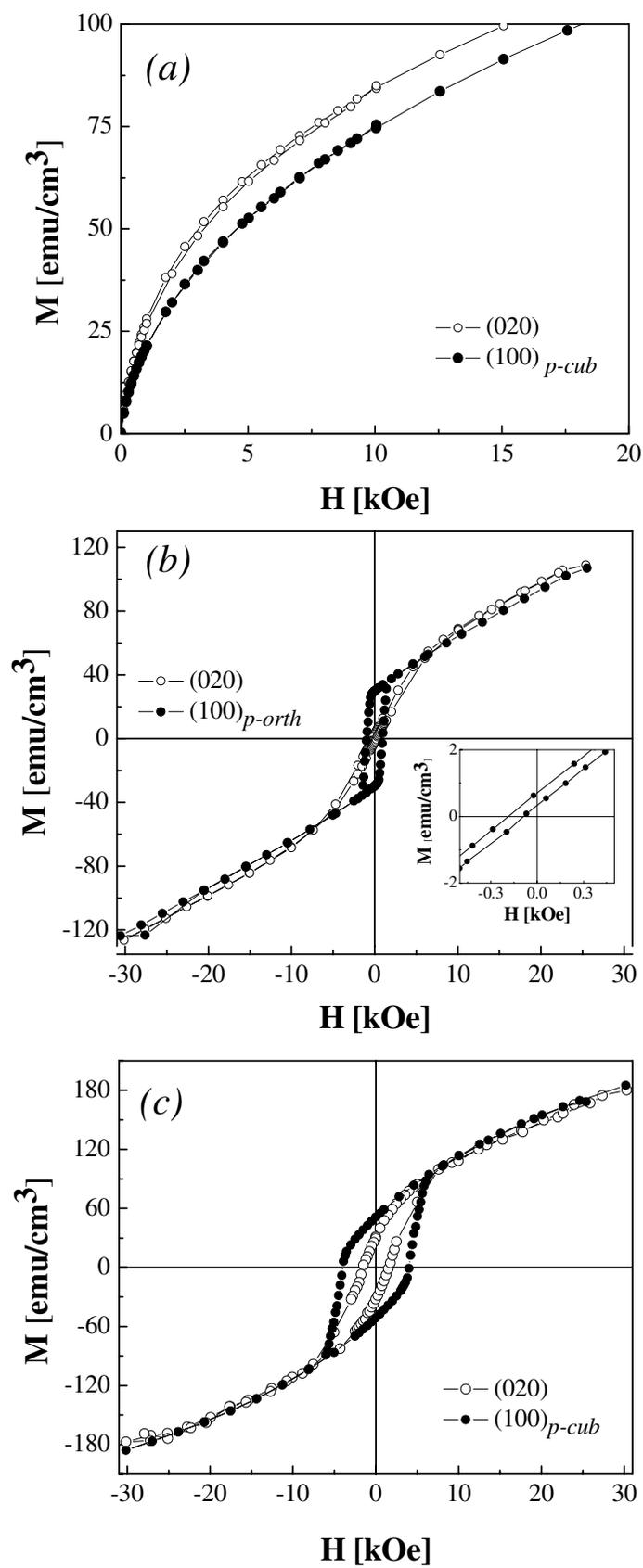

**Figure**

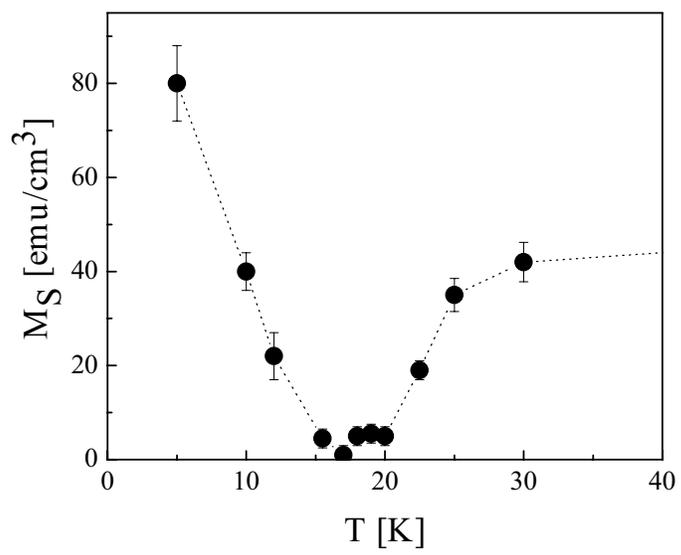

**Figure**

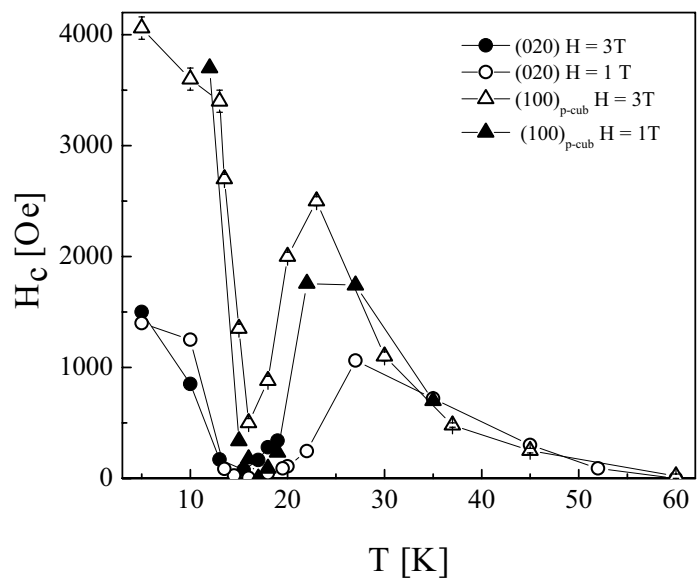

**Figure**

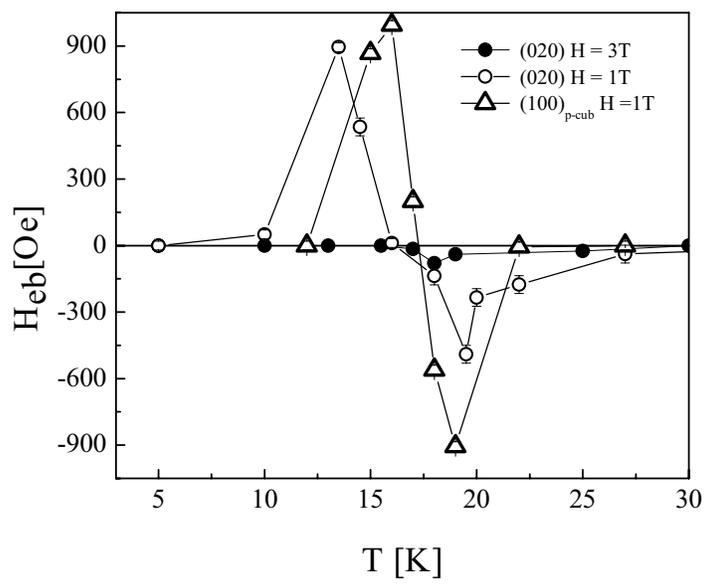